\documentclass[final,5p,times]{elsarticle}
\usepackage{amsfonts}
\usepackage{amssymb,bbold}
\usepackage{amsmath}
\usepackage{float}
\usepackage{natbib}
\usepackage{graphicx}
\usepackage{hyperref}

\makeatletter
\newcommand{\bea}{\begin{eqnarray}}
\newcommand{\eea}{\end{eqnarray}}

\hypersetup{
  colorlinks=true,linkcolor=blue,citecolor=blue,
  filecolor=blue,urlcolor=blue,breaklinks=true
}
\biboptions{sort&compress}
\journal{Physics Letters B}
\begin{document}

\date{\today}

\begin{frontmatter}

\title{Symplectic Dirac Equation}

\author[unb,gama]{R. G. G. Amorim}
\ead{ronniamorim@gmail.com}

\author[unb]{S. C. Ulhoa}
\ead{sc.ulhoa@gmail.com}

\author[ufma]{Edilberto O. Silva}
\ead{edilbertoo@gmail.com}

\address[unb]{
  Instituto de F\'{i}sica,
  Universidade de Bras\'{i}lia,
  70910-900, Bras\'{i}lia, Distrito Federal, Brazil
}
\address[gama]{Faculdade Gama, Universidade de Bras\'{\i}lia, Setor Leste (Gama),
72444-240, Bras\'{\i}lia, Distrito Federal, Brazil}
\address[ufma]{Departamento de F\'{i}sica, Universidade Federal do Maranh\~{a}o,
  Campus Universit\'{a}rio do Bacanga, 65085-580, S\~{a}o Lu\'{i}s, Maranh\~{a}o, Brazil
}

\begin{abstract}
Symplectic unitary representations for the Poincar\'{e} group are studied.
The formalism is based on the noncommutative structure of the star-product,
and using group theory approach as a guide, a consistent physical theory in
phase space is constructed. The state of a quantum mechanics system is
described by a quasi-probability amplitude that is in association with the
Wigner function. As a result, the Klein-Gordon and Dirac equations are
derived in phase space. As an application, we study the Dirac equation with
electromagnetic interaction in phase space.
\end{abstract}


\end{frontmatter}

\section{Introduction}

The first formalism to quantum mechanics in phase space was proposed by
Wigner in 1932 \cite{wig1}. He was motived by the problem of finding a way
to improve the quantum statistical mechanics, based on the desity matrix, to
treat the transport equations for superfluids \cite{wig2, wig3, wig4}. Since
then, the formalism proposed by Wigner has been applied in different
contexts, such as quantum optics \cite{zak1, ViannaCezar}, condensed matter \cite{seb4, seb41, seb42}, quantum computing \cite{seb43, seb44, seb445},
quantum tomography \cite{dav02}, plasma physics\cite{seb5, seb6, seb7, seb9, seb10, seb12}.
Wigner introduced his formalism by using a kind of Fourier transform of the
density matrix, $\rho (q,q^{\prime })$, giving rise to what in nowadays
called the Wigner function, $f_{W}(q,p)$, where $(q,p)$ are coordinates of a
phase space manifold $(\Gamma )$. The Wigner function is identified as a
quasi-probability density in the sense that $f_{W}(q,p)$ is real but not
positive defined, and as such cannot be interpreted as a probability.
However, the integrals $\sigma (q)=\int f_{W}(q,p)dp$ and $\sigma (p)=\int
f_{W}(q,p)dq$ are distribution functions \cite{wig1, wig2}.

In Wigner formalism each quantum operator $A$ in the Hilbert space is
associated with a function $a_{W}(q,p)$ defined in $\Gamma $. The
application $\Omega _{W}:A\rightarrow a_{W}(q,p)$ is such that associative
algebra of operators in $\mathcal{H}$ defines an associative but
noncommutative algebra in $\Gamma $. The noncommutativity stems from nature
of the product between two operators in $\mathcal{H}$. Given two operators $%
A $ and $B$, we have the mapping $\Omega :AB\rightarrow a_{W}(q,p)\star
b_{W}(q,p)$, where the star (or Moyal)-product $\star $ is defined by \cite%
{moy2}
\begin{equation}
a_{w}(q,p)\star b_{w}(q,p)=a_{w}(q,p)\exp [\frac{i}{2}(\frac{\overleftarrow{
\partial }}{\partial q}\frac{\overrightarrow{\partial }}{\partial p}-\frac{%
\overleftarrow{\partial }}{\partial p}\frac{\overrightarrow{\partial }}{%
\partial q})]b_{w}(q,p).  \label{1}
\end{equation}%
(Throught this Letter we use natural units: $\hbar =c=1$). Note that Eq.(\ref%
{1}) can be seen as an operator $\hat{A}=a_{W}(q,p)\star $ acting on
functions $b_{W}(q,p)$, such that $\hat{A}(b_{W})=a_{W}\star b_{W}$. In this
sense, we can study unitary representations of Lie groups in phase space
using the Moyal product as defined by the operators $\hat{A}$. This gives
rise, for instance, to the Klein-Gordon and Dirac equations written in phase
space \cite{seb1, seb2, seb22, sig1}. The connection with Wigner function is
derived, providing a physical interpretation for the representation. As a
consequence, these symplectic representations are a a way to consider the
Wigner methods on the bases of symmetry groups. In the present work, we
apply this symplectic formalism to solve Dirac equation with electromagnetic
interaction in phase space. These results provide a starting point for our
analysis of nonclassical electromagnetic radiation sates in phase space.

The presentation of this Letter is organized in the following way. In
section 2, we define a Hilbert space $\mathcal{H}(\Gamma)$ over a phase
space $\Gamma$ with its natural relativistic symplectic struture. In section
3, we study the Poincar\'{e} algebra in $\mathcal{H}(\Gamma)$ and the
representation for spin 1/2. In section 4, the Dirac equation in phase space
with electromagnetic radiation is considered. Quasi-amplitudes of
probabilities are derived . In section 5, final concluding remarks are
presented.

\section{Hilbert Space and Symplectic Structure}

Consider $M$ an $n$-dimensional analytical manifold where each point is
specified by Minkowski coordinates $q^{\mu}$, with $\mu=0,1,2,3,4$ and
metric specified by $diag(g)=(-+++)$. The coordinates of each point in $%
T^{*}M$ will be denoted by $(q^{\mu},p^{\mu})$. The space $T^{*}M$ is
equipped with a symplectic struture by introducing a 2-form
\begin{equation}
\omega =dq^{\mu}\wedge dp_{\mu},
\end{equation}
called the symplectic form (sum over repeated indices is assumed). Consider
the following bidifferential operator on $C^{\infty}(T^{\ast}M)$:
\begin{equation}
\Lambda=\frac{\overleftarrow{\partial}}{\partial q^{\mu}}\frac{%
\overrightarrow{\partial}}{\partial p_{\mu}}-\frac{\overleftarrow{\partial}}{%
\partial p^{\mu}}\frac{\overrightarrow{\partial}}{\partial q_{\mu}},
\end{equation}
such that for $C^{\infty}$ functions, $f=f(q^{\mu},p^{\mu})$ and $%
g=g(q^{\mu},p^{\mu})$, we have
\begin{equation}
\{f,g\}=\omega(f\Lambda,g\Lambda)=f\Lambda g,
\end{equation}
where
\begin{equation}
\{f,g\}=\frac{\partial f}{\partial q^{\mu}}\frac{\partial g}{\partial p_{\mu}%
}-\frac{\partial f}{\partial p^{\mu}}\frac{\partial g}{\partial q_{\mu}}.
\end{equation}
is the Poisson bracket and $f\Lambda$ and $g\Lambda$ are two vector fields
given by $h\Lambda=X_{h}=-\{h,\}$. The space $T^{\ast}$
endowed with this symplectic structure is called the phase space, and will
be denoted by $\Gamma$.

The notion of Hilbert space associated with the phase space $\Gamma $ is
introduced by considering the set of square integrable functions, $\phi
(q^{\mu },p^{\mu })$ in $\Gamma $, such that
\begin{equation}
\int dp^{\mu }dq^{\mu }\phi ^{\ast }(q^{\mu },p^{\mu })\phi (q^{\mu },p^{\mu
})<\infty .
\end{equation}%
Then, we can write $\phi (q^{\mu },p^{\mu })=\langle q^{\mu },p^{\mu }|\phi
\rangle $, with
\begin{equation}
\int dp^{\mu }dq^{\mu }|q^{\mu },p^{\mu }\rangle \langle q^{\mu },p^{\mu
}|=1,
\end{equation}%
to be $\langle \phi |$ the dual vector of $|\phi \rangle $. We call this the
Hilbert space $\mathcal{H}(\Gamma )$.

\section{Poincar\'{e} Algebra and Dirac Equation in Phase Space}

Using the $star$-operators, $\hat{A}=a_{W}(q,p)\star $, we define 4-momentum
and 4-position operators, respectively, by
\begin{equation}
\hat{P}^{\mu }=p^{\mu }\star =p^{\mu }\exp [\frac{i}{2}(\frac{\overleftarrow{%
\partial }}{\partial q^{\mu }}\frac{\overrightarrow{\partial }}{\partial
p_{\mu }}-\frac{\overleftarrow{\partial }}{\partial p^{\mu }}\frac{%
\overrightarrow{\partial }}{\partial q_{\mu }})]=p^{\mu }-\frac{i}{2}\frac{%
\partial }{\partial q_{\mu }},  \label{mom}
\end{equation}%
\begin{equation}
\hat{Q}^{\mu }=q^{\mu }\star =q^{\mu }\exp [\frac{i}{2}(\frac{\overleftarrow{%
\partial }}{\partial q^{\mu }}\frac{\overrightarrow{\partial }}{\partial
p_{\mu }}-\frac{\overleftarrow{\partial }}{\partial p^{\mu }}\frac{%
\overrightarrow{\partial }}{\partial q_{\mu }})]=q^{\mu }+\frac{i}{2}\frac{%
\partial }{\partial p_{\mu }}.  \label{pos}
\end{equation}%
From Eqs. (\ref{mom}) and (\ref{pos}), we can introduce the quantity $\hat{M}%
_{\nu \sigma }=\hat{Q}_{\mu }\hat{P}_{\sigma }-\hat{Q}_{\sigma }\hat{P}_{\mu
}$. The operators $\hat{P}^{\mu }$ and $\hat{M}_{\nu \sigma }$ are defined
in the Hilbert space, $\mathcal{H}(\Gamma )$, constructed with complex
functions in the phase space $\Gamma $, and satisfy the set of commutation
relations
\begin{equation}
\lbrack \hat{M}_{\mu \nu },\hat{P}_{\sigma }]=i(g_{\nu ,\sigma }\hat{P}_{\mu
}-g_{\sigma \mu }\hat{P}_{\nu }),
\end{equation}%
\begin{equation}
\lbrack \hat{P}_{\mu },\hat{P}_{\nu }]=0,
\end{equation}%
\begin{equation}
\lbrack \hat{M}_{\mu \nu },\hat{M}_{\sigma \rho }]=-i(g_{\mu \rho }\hat{M}%
_{\nu \sigma }-g_{\nu \rho }\hat{M}_{\mu \sigma }+g_{\mu \sigma }\hat{M}%
_{\rho \nu }-g_{\nu \sigma }\hat{M}_{\rho \mu }).
\end{equation}%
This is the Poincar\'{e} algebra, where $\hat{M}_{\mu \nu }$ stands for
rotations and $\hat{P}_{\mu }$ for translations (but notice, in phase
space). The Casimir invariants are calculated by using the Pauli-Lubanski
matrices, $\hat{W}_{\mu }=\frac{1}{2}\epsilon _{\mu \nu \rho \sigma }\hat{M}%
^{\mu \sigma }\hat{P}^{\rho }$, where $\epsilon _{\mu \nu \rho \sigma }$ is
the Levi-Civita symbol. The invariants are
\begin{equation}
\hat{P}^{2}=\hat{P}^{\mu }\hat{P}_{\mu },
\end{equation}%
and
\begin{equation}
\hat{W}^{2}=\hat{W}^{\mu }\hat{W}_{\mu },
\end{equation}%
where $\hat{P}^{2}$ stands for the mass shell condition and $\hat{W}^{2}$
for the spin.

To determine the Klein-Gordon field equation, we consider a scalar
representation in $\mathcal{H}(\Gamma )$. In this case, we can use the
invariant $\hat{P}^{2}$ to write
\begin{eqnarray}
\hat{P}^{2}\phi (q^{\mu },p^{\mu }) &=&(p^{2})\star \phi (q^{\mu },p^{\mu }),
\notag \\
&=&(p^{\mu }\star p_{\mu }\star )\phi (q^{\mu },p^{\mu }),  \notag \\
&=&m^{2}\phi (q^{\mu },p^{\mu }),  \label{kg1}
\end{eqnarray}%
where $m$ is a constant fixing the representation and interpreted as mass,
such that the mass shell condition is satisfied. Using Eq. (\ref{mom}), we
obtain
\begin{equation}
\left( p^{\mu }p_{\mu }-ip^{\mu }\frac{\partial }{\partial q^{\mu }}-\frac{1%
}{4}\frac{\partial }{\partial q^{\mu }}\frac{\partial }{\partial q_{\mu }}%
\right) \phi (q^{\mu },p^{\mu })=m^{2}\phi (q^{\mu },p^{\mu }),  \label{kg2}
\end{equation}%
which is the Klein-Gordon equation in phase space.

The association of this representation with Wigner formalism is given by
\cite{seb2}
\begin{equation}  \label{w1}
f_W(q^{\mu},p^{\mu})=\phi(q^{\mu},p^{\mu})\star\psi^{\star}(q^{\mu},p^{\mu}),
\end{equation}
where $f_W(q^{\mu},p^{\mu})$ is the relativistic Wigner function.

The representation for spin-$1/2$ leads to
\begin{equation}  \label{d1}
\gamma^{\mu}\left(p_{\mu}-\frac{i}{2}\frac{\partial}{\partial q^{\mu}}%
\right)\psi(q^{\mu},p^{\mu})=m\psi(q^{\mu},p^{\mu}),
\end{equation}
which is the Dirac equation in phase space, where the $\gamma^{\mu}$%
-matrices fulfill the usual Clifford algebra, $(\gamma^{\mu}\gamma^{\nu}-%
\gamma^{\nu}\gamma^{\mu})=2g^{\mu\nu}$. The Wigner function, in this case,
is given by \cite{seb2}
\begin{equation}  \label{w2}
f_W(q^{\mu},p^{\mu})=\psi(q^{\mu},p^{\mu})\star\overline{\psi}%
(q^{\mu},p^{\mu}),
\end{equation}
where $\overline{\psi}(q^{\mu},p^{\mu})=\gamma^0\psi^{\dagger}(q^{\mu},p^{%
\mu})$, with $\psi^{\dagger}(q^{\mu},p^{\mu})$ being the Hermitian of $%
\psi(q^{\mu},p^{\mu})$. We point out that the CPT theorem holds for non-commutative theories as showed in \cite{ch}. Therefore, such a theorem is also valid in phase space since the group structure remains the same.

One central point to be emphasized is that the approach developed here permits the calculation of Wigner functions for relativistic systems with methods,
based on symmetry, similar to those used in quantum field theory.

\section{Solution of Dirac Equation with Electromagnetic Interaction on
Phase Space}

In this section, we study interactions of a spin-1/2 charged particle with
an external electromagnetic field in Phase Space. The relevant equation is
the Dirac equation with minimal coupling

\begin{equation}
\left( \gamma ^{\mu }\hat{P}_{\mu }+m\right) \Psi =0,
\end{equation}%
being
\begin{equation}
\hat{P}_{\mu }\rightarrow \hat{P}_{\mu }-e\hat{A}_{\mu },
\end{equation}%
the minimal coupling prescription, where $\hat{A}^{i}=\frac{1}{2}\epsilon
^{ijk}B_{j}\hat{X}_{k}$ and $\hat{A}^{0}=0$, which represents the chosen
gauge. We also chose the magnetic field as $\mathbf{B}=(0,0,B)$. Thus, we
have
\begin{equation}
\left[ \gamma ^{\mu }\left( \hat{P}_{\mu }-e\hat{A}_{\mu }\right) +m\right]
\Psi =0.  \label{edirac}
\end{equation}%
Now, we make the definition
\begin{equation}
\Psi =\left[ \gamma ^{\mu }\left( \hat{P}_{\mu }-e\hat{A}_{\mu }\right) -m%
\right] \psi =0.  \label{defsol}
\end{equation}%
In order to obtain the energy levels, we substitute Eq. (\ref{defsol}) into
Eq. (\ref{edirac}) to give
\begin{equation}
\left[ \gamma ^{\mu }\gamma ^{\nu }\left( \hat{P}_{\mu }-e\hat{A}_{\mu
}\right) \left( \hat{P}_{\nu }-e\hat{A}_{\nu }\right) -m^{2}\right] \psi =0,
\label{ediracc}
\end{equation}%
with
\begin{equation}
\gamma ^{\mu }\gamma ^{\nu }=g^{\mu \nu }+\sigma ^{\mu \nu }\,,
\end{equation}%
where%
\begin{equation}
\sigma ^{\mu \nu }=\frac{i}{2}(\gamma ^{\mu }\gamma ^{\nu }-\gamma ^{\nu
}\gamma ^{\mu })=\frac{i}{2}[\gamma ^{\mu },\gamma ^{\nu }].  \label{sigma}
\end{equation}%
The components $\sigma ^{0i}$, $\sigma ^{ij}$ of the operator (\ref{sigma})
are \
\begin{equation}
\sigma ^{0i}=i\left(
\begin{array}{cc}
0 & \sigma ^{i} \\
\sigma ^{i} & 0%
\end{array}%
\right) ,\text{ \ \ }\sigma ^{ij}=-\left(
\begin{array}{cc}
\epsilon _{ijk}\sigma ^{k} & 0 \\
0 & \epsilon _{ijk}\sigma ^{k}%
\end{array}%
\right) .
\end{equation}%
Note that these components are also expressed as $\sigma ^{0j}=i\alpha ^{j},$
$\sigma ^{ij}=-\epsilon _{ijk}\Sigma ^{k}.$ These results are explicitly
evaluated in the following representation of the $\gamma $-matrices:
\begin{align}
\gamma ^{0}& =\left(
\begin{array}{cc}
1 & 0 \\
0 & -1%
\end{array}%
\right) ,\;\;\;\gamma ^{i}=\left(
\begin{array}{cc}
0 & \sigma ^{i} \\
-\sigma ^{i} & 0%
\end{array}%
\right) ,\;\;\;\gamma _{5}=\left(
\begin{array}{cc}
0 & 1 \\
1 & 0%
\end{array}%
\right) ,  \notag \\
\alpha ^{i}& =\left(
\begin{array}{cc}
0 & \sigma ^{i} \\
\sigma ^{i} & 0%
\end{array}%
\right) ,\;\;\;\Sigma ^{k}=\left(
\begin{array}{cc}
\sigma ^{k} & 0 \\
0 & \sigma ^{k}%
\end{array}%
\right) .
\end{align}%
with $\mathbf{\sigma }=(\sigma _{x},\sigma _{y},\sigma _{z})$ being the
Pauli matrices. Equation (\ref{ediracc}) can be written as
\begin{equation}
\left\{ \hat{P}^{\mu }\hat{P}_{\mu }-e\left( \hat{P}^{\mu }\hat{A}_{\mu }+%
\hat{A}^{\mu }\hat{P}_{\mu }\right) +e^{2}\hat{A}^{\mu }\hat{A}_{\mu
}+e\sigma ^{\mu \nu }\left[ \hat{P}^{\nu },\hat{A}_{\mu }\right]
-m^{2}\right\} \psi =0.
\end{equation}%
Confining the motion at the plane $\hat{X}\hat{Y}$ by the choice $\hat{P}%
_{3}=0$ and using the operators $\hat{P}_{\mu }=p_{\mu }-\frac{i}{2}\frac{%
\partial }{\partial x^{\mu }}$ and $\hat{X}_{\mu }=x_{\mu }-\frac{i}{2}\frac{%
\partial }{\partial p^{\mu }}$, we get the following equation
\begin{align}
& \left( -E^{2}-iE\frac{\partial }{\partial t}+\frac{1}{4}\frac{\partial ^{2}%
}{\partial t^{2}}-m^{2}\right) \psi +\Biggl\{p_{x}^{2}+p_{y}^{2}-\frac{1}{4}%
\left( \frac{\partial ^{2}}{\partial x^{2}}+\frac{\partial ^{2}}{\partial
y^{2}}\right)   \notag \\
& -eB\left[ \frac{i}{2}\left( p_{y}\frac{\partial }{\partial p_{x}}-p_{x}%
\frac{\partial }{\partial p_{y}}\right) +\frac{1}{4}\left( \frac{\partial
^{2}}{\partial y\partial p_{x}}-\frac{\partial ^{2}}{\partial x\partial p_{y}%
}\right) \right]   \notag \\
& -i\left( p_{y}\frac{\partial }{\partial y}-p_{x}\frac{\partial }{\partial x%
}\right) -eB\left[ \left( xp_{y}-yp_{x}\right) -\frac{i}{2}\left( x\frac{%
\partial }{\partial y}-y\frac{\partial }{\partial x}\right) \right]   \notag
\\
& +\frac{e^{2}B^{2}}{4}\left[ \left( x+\frac{i}{2}\frac{\partial }{\partial
p_{x}}\right) ^{2}+\left( y+\frac{i}{2}\frac{\partial }{\partial p_{y}}%
\right) ^{2}\right] +ieB\sigma ^{12}\Biggr\}\psi =0.  \label{dracph}
\end{align}%
Since $\gamma _{5}$ commutes through all terms of Eq. (\ref{dracph}) and, if
we have found the solution $\psi $, then we must also have that $\gamma
_{5}\psi $ is a solution. In this case, In this case, the wave function can
takes on one of the forms%
\begin{equation}
\psi =\left(
\begin{array}{c}
\psi _{1} \\
\psi _{2} \\
\psi _{3} \\
\psi _{4}%
\end{array}%
\right) ,~~~\psi =\left(
\begin{array}{c}
\psi _{1} \\
\psi _{2} \\
-\psi _{1} \\
-\psi _{2}%
\end{array}%
\right) .  \label{newsol}
\end{equation}%
Thus, we can select only one of these solutions as that will make the other
redundant. We specialize the solution
\begin{equation}
\gamma _{5}\psi =\psi .  \label{newsolb}
\end{equation}%
We can write $\psi $ in terms of $\Psi $ in Eq. (\ref{defsol}) as
\begin{equation}
\psi =\frac{1}{2}\left( I+\gamma _{5}\right) \Psi ,  \label{solp}
\end{equation}%
where $I$ is the unit matrix. From Eq. (\ref{newsol}), we can show that Eq. (%
\ref{solp}) can be put in the form%
\begin{equation}
\psi =\left(
\begin{array}{c}
\chi (E,t,p_{x},p_{y},x,y) \\
-\chi (E,t,p_{x},p_{y},x,y)%
\end{array}%
\right) =\left(
\begin{array}{c}
\varphi (E,t)\phi \left( p_{x},p_{y},x,y\right)  \\
-\varphi (E,t)\phi \left( p_{x},p_{y},x,y\right)
\end{array}%
\right) ,  \label{soltrue}
\end{equation}%
where $\chi (E,t,p_{x},p_{y},x,y)$ is a two-component wavefunction. Note
that, in the representation (\ref{soltrue}), the upper two components of Eq.
(\ref{dracph}) are now completely decoupled from the lower two. So, we have,
in two-component form, the following equations:
\begin{equation}
\left( -E^{2}-iE\frac{\partial }{\partial t}+\frac{1}{4}\frac{\partial ^{2}}{%
\partial t^{2}}-m^{2}\right) \varphi =\lambda ^{2}\varphi ,  \label{solE}
\end{equation}%
\begin{align}
& \Biggl\{p_{x}^{2}+p_{y}^{2}-\frac{1}{4}\left( \frac{\partial ^{2}}{%
\partial x^{2}}+\frac{\partial ^{2}}{\partial y^{2}}\right) -i\left( p_{y}%
\frac{\partial }{\partial y}-p_{x}\frac{\partial }{\partial x}\right)
\notag \\
& -eB\Big[\left( xp_{y}-yp_{x}\right) +\frac{i}{2}\left( p_{y}\frac{\partial
}{\partial p_{x}}-p_{x}\frac{\partial }{\partial p_{y}}\right)   \notag \\
& -\frac{i}{2}\left( x\frac{\partial }{\partial y}-y\frac{\partial }{%
\partial x}\right) +\frac{1}{4}\left( \frac{\partial ^{2}}{\partial
y\partial p_{x}}-\frac{\partial ^{2}}{\partial x\partial p_{y}}\right) \Big]
\notag \\
& +\frac{e^{2}B^{2}}{4}\left[ \left( x+\frac{i}{2}\frac{\partial }{\partial
p_{x}}\right) ^{2}+\left( y+\frac{i}{2}\frac{\partial }{\partial p_{y}}%
\right) ^{2}\right] +ieB\sigma ^{12}\Biggr\}\phi =\lambda ^{2}\phi ,
\label{solxp}
\end{align}%
where $\lambda$ is a constant. We point out that $E$ is not associated to $i\frac{\partial }{\partial t}$ a priori in Eq. \ref{solE}, thus the energy comes from $\lambda$. Note that Eqs. (\ref{solE}) and (\ref{solxp}%
) determines $\chi $ and, hence, from Eq. (\ref{solp}), it determines $\psi $%
. Performing a changing of variables in Eq. (\ref{solxp}) of the form
\begin{equation*}
z=p_{x}^{2}+p_{y}^{2}+eB\left( yp_{x}-xp_{y}\right) +\frac{e^{2}B^{2}}{4}%
\left( x^{2}+y^{2}\right) ,
\end{equation*}%
we note that the imaginary part of the equation vanished which yields
\begin{equation}
z\phi -e^{2}B^{2}\dot{\phi}-e^{2}B^{2}z\ddot{\phi}=\left( \lambda
^{2}+seB\right) \phi ,
\end{equation}%
where we have used $i\sigma ^{12}\phi =-s\phi $, with $s=\pm 1$. If we use $%
\omega =z/eB$ and $\phi =\exp {(-\omega )}F(\omega )$, the equation for $%
F(\omega )$ is found to be
\begin{equation}
\omega F^{^{\prime \prime }}+(1-2\omega )F^{^{\prime }}-(1-k)F=0,
\label{hyper}
\end{equation}%
where $F^{\prime }\equiv \frac{dF}{d\omega }$ and $k=(\lambda ^{2}+seB)/eB$.
Equation (\ref{hyper}) is of the confluent hypergeometric equation type
\begin{equation}
zF^{\prime \prime }(z)+(b-z)F^{\prime }(z)-aF(z)=0.
\end{equation}%
In this manner, the general solution for Eq. (\ref{hyper}) is given by
\begin{equation}
F\left( \omega \right) =\mathit{A}_{\mathit{m}}\,{\mathrm{M}\left( \frac{1}{2%
}-\frac{\,k}{2},\,1,\,2\,\omega \right) }+\mathit{B}_{\mathit{m}}\,{\mathrm{U%
}\left( \frac{1}{2}-\frac{\,k}{2},\,1,\,2\,\omega \right) ,}  \label{kummer}
\end{equation}%
where $\mathrm{M}(a,b,z)$, $\mathrm{U}(a,b,z)$ are the Kummer functions, and
$\mathit{A}_{\mathit{m}}$, $\mathit{B}_{\mathit{m}}$ are constants. Since
only $\mathrm{U}(a,b,z)$ is square integrable, we consider it as a physical
solution. Thus, we can impose that $\mathit{A}_{\mathit{m}}=0$. Furthermore,
if $a=-n$, with $n=0,1,2,\ldots ,$ the series $\mathrm{U}(a,b,z)$ becomes a
polynomial in $z$ of degree not exceeding $n$. From this condition, we can
write%
\begin{equation}
1-k=-2n,  \label{cpd}
\end{equation}%
from which, we can extract the relation%
\begin{equation}
\lambda ^{2}=eB\left( 2n+1+s\right) .  \label{engy}
\end{equation}%
The wave function is given by
\begin{equation}
f_{m}(z)=\mathit{B}_{\mathit{m}}\,\,\mathrm{M}\,{\left( -n,\,1,\,\frac{2\,z}{eB%
}\right) },  \label{wf1}
\end{equation}%
where $\mathit{B}_{\mathit{m}}\,$\ is a normalization constant.

The Wigner function related to Dirac equation with an electromagnetic interaction is formally given by
$$
f_W(x,y,p_x,p_y)=\Psi_n(x,y,p_x,p_y)\star\overline{\Psi}_n(x,y,p_x,p_y).
$$
Thus Wigner function is used to determine mean values, for example, this result can be useful for theoretical and applied areas, such as: quantum optics, quantum tomography and quantum computing. We point out that Landau levels which appear in expression (\ref{engy}) represent as a matter of fact a planar oscillator and the variable z in Eq. (\ref{wf1}) give us information about the symplectic structure.

\section{Conclusion}

We have set forth a symplectic representation of the Poincar\'{e} group,
which yields quantum theories in phase space. We have derived the
Klein-Gordon and Dirac equations in phase space and, as illustrations,
studied the Dirac equation with electromagnetic interaction. The symplectic
representation is constructed on the basis of the Moyal or star product, an
ingredient of noncommutative geometry. A Hilbert space is then defined from
a manifold with the features of phase space. The states are represented by a
quasi-amplitude of probability, a wave function in phase space, the
definition of which makes connection with the Wigner function, i.e., the
quasi-probability density. Nontrivial, yet consistent, the association with
the Wigner function provides a physical interpretation of the theory.
Analogous interpretations are not found in other studies of representations
in phase space [25, 26]. One aspect of the procedure deserves emphasis. Our
formalism explores unitary representations to calculate Wigner functions.
This constitutes an important advantage over the more traditional
constructions of the Wigner method, which entail several intricacies
associated with the Liouville-von Neumann equation.  Furthermore, the
formalism we have described opens new perspectives for applications of the
Wigner function method in quantum field theory. This aspect of the formalism
will be discussed in a forthcoming paper.

\section*{Acknowledgments}

This work was supported by the CNPq, Brazil, Grants No. 482015/2013-6 (Universal) and No. 306068/2013-3 (PQ); FAPEMA, Brazil, Grants No. 00845/13 (Universal) and No. 01852/14 (PRONEM).


\end{document}